# Stable surface solitons in truncated complex potentials

Yingji He,[1,*] Dumitru Mihalache,[2] Xing Zhu,[3] Lina Guo,[1] and Yaroslav V. Kartashov[4]

[1]*School of Electronics and Information, Guangdong Polytechnic Normal University, 510665 Guangzhou, China*
[2]*Horia Hulubei National Institute for Physics and Nuclear Engineering, P.O.B. MG-6, 077125 Magurele-Bucharest, Romania*
[3]*State Key Laboratory of Optoelectronic Materials and Technologies, Sun Yat-Sen University, 510275 Guangzhou, China*
[4]*Institute of Spectroscopy, Russian Academy of Sciences, Troitsk, Moscow Region 142190, Russia*
*\*Corresponding author: heyingji8@126.com*

We show that surface solitons in the one-dimensional nonlinear Schrödinger equation with truncated complex periodic potential can be stabilized by linear homogeneous losses, which are necessary to balance gain in the near-surface channel arising from the imaginary part of potential. Such solitons become stable attractors when the strength of homogeneous losses acquires values from a limited interval and they exist in focusing and defocusing media. The domains of stability of surface solitons shrink with increase of the amplitude of imaginary part of complex potential. © 2012 Optical Society of America

The evolution of nonlinear excitations in complex-valued external potentials attracts considerable attention. A separate place among such potentials is occupied by $\mathcal{PT}$-symmetric periodic structures that may support stable soliton families despite periodic alternation of spatial domains with linear [1-4] or nonlinear [5-7] gain and losses. The truncation of such potentials breaks their symmetry and prohibits the formation of stationary states, at least around the edge of potential. Another interesting class of complex potentials is represented by structures where linear gain acts only in one or several localized spots, while real part of potential may be periodic. Such potentials support dissipative solitons if additional nonlinear losses compensate localized gain [8-11]. Notice that in sharp contrast to $\mathcal{PT}$-symmetric structures the truncation of such potentials still allows existence of stationary states even around the edge of potential if gain also acts around its edge [12].

Surface solitons [13,14] that may form at the edge of truncated periodic lattices were first thoroughly investigated in conservative systems. Such states were suggested in [15] and subsequently observed in one- and two-dimensional settings, in both focusing and defocusing media [16-21]. Distinctive feature of surface solitons is the existence of threshold power required for their formation that can be made relatively low in potentials with shallow refractive index modulation. At the same time surface solitons in complex potentials were studied only in settings where gain arising from imaginary part of potential is compensated by nonlinear losses [12], i.e. only in systems where several dissipative mechanisms compete with each other, as in Ginzburg-Landau equation [22]. Stable surface states in settings with only linear gain and losses were not reported so far.

In this Letter we show that such solitons may form at the edge of truncated periodic complex potential in the presence of homogeneous losses in the region occupied by potential. Such solitons are stable attractors and they can form in both focusing and defocusing media.

We describe the propagation of light beam along the interface of complex-valued potential with the nonlinear Schrödinger equation for the light field amplitude $q$:

$$i\frac{\partial q}{\partial \xi} = -\frac{1}{2}\frac{\partial^2 q}{\partial \eta^2} + \sigma q|q|^2 - \{p_r R_r(\eta) - i[p_i R_i(\eta) + \gamma]\}H(\eta)q, \quad (1)$$

where $\eta, \xi$ are the normalized transverse and longitudinal coordinates, respectively; $\sigma = \mp 1$ corresponds to focusing/defocusing nonlinearity; $p_r, p_i$ are the amplitudes of real and imaginary parts of potential; $\gamma \leq 0$ is the coefficient of losses; $H(\eta) = 0$ for $\eta < 0$ and $H(\eta) = 1$ for $\eta \geq 0$; the functions $R_r(\eta) = \cos^2(\Omega\eta)$ and $R_i(\eta) = \sin(2\Omega\eta)$ describe real and imaginary parts of truncated potential. Further we set $p_r = 4$, $\Omega = 1$ and vary $p_i$ and $\gamma$.

For $\gamma = 0$ and without truncation complex potential $p_r R_r(\eta) - ip_i R_i(\eta)$ is $\mathcal{PT}$-symmetric. Thus, it can support soliton families $q(\eta,\xi) = [w_r(\eta) + iw_i(\eta)]\exp(ib\xi)$, whose properties are determined by the propagation constant $b$, like in conservative systems [1,2]. The truncation of potential breaks its symmetry and stationary solitons cannot be found anymore in the vicinity of the edge of potential. The central result of this Letter is that such solutions become possible in the presence of homogeneous losses acting in the region $\eta \geq 0$. Such solitons are truly dissipative and they become stable attractors (in contrast to solitons in $\mathcal{PT}$-symmetric lattices) for certain values of $\gamma$. In order to find them we used Newton method with zero boundary conditions for the field $w_{r,i}|_{\eta \to \pm\infty} = 0$, supplied by the energy balance condition, which is necessary since $b$ is not a free parameter like in standard approaches for calculation of soliton shapes in conservative media [23]. To analyze their stability we write perturbed soliton solutions as $q(\eta,\xi) = [w_r + iw_i + (u+iv)\exp(\delta\xi)]\exp(ib\xi)$, where $u,v$ are real and imaginary parts of small perturbation satisfying $u,v|_{\eta\to\pm\infty} = 0$, substitute them into Eq. (1) and obtain the linear eigenvalue problem:

$$\begin{aligned}
\delta u &= -\frac{1}{2}\frac{d^2 v}{d\eta^2} - p_r H R_r v + H(p_i R_i + \gamma)u + bv + \\
&\quad 2\sigma u w_i w_r + \sigma v(3w_i^2 + w_r^2), \\
\delta v &= +\frac{1}{2}\frac{d^2 u}{d\eta^2} + p_r H R_r u + H(p_i R_i + \gamma)v - bu - \\
&\quad \sigma u(3w_r^2 + w_i^2) - 2\sigma v w_i w_r,
\end{aligned} \quad (2)$$

that was solved with standard eigenvalue solver providing eigenvalues (perturbation growth rates $\delta$) for various $\gamma$.

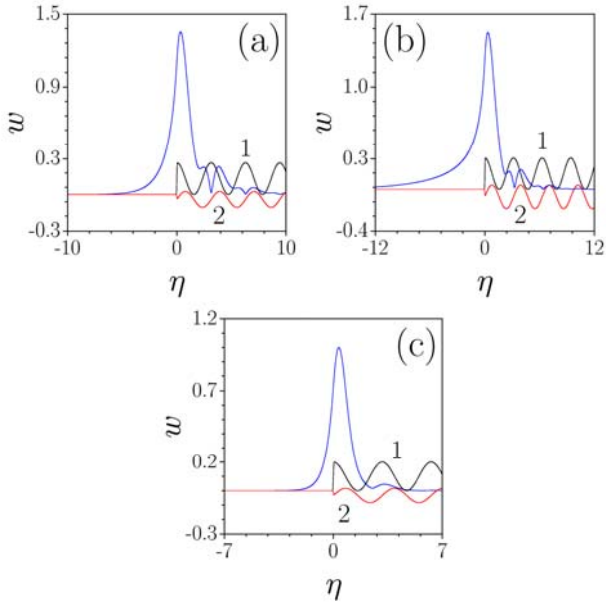

Fig. 1. (Colour online) Field modulus distribution (blue curves) for surface solitons in defocusing medium at $\gamma = -0.6224$ and $p_i = 1$ (a) and $\gamma = -0.9590$ at $p_i = 1.5$ (b). Field modulus distribution for surface soliton in focusing medium at $\gamma = -0.6671$ and $p_i = 1$ (c). Real $p_r R_r$ (curves 1) and imaginary $p_i R_i + \gamma$ (curves 2) parts of potential were downscaled for illustrative purposes.

The representative examples of surface solitons are shown in Fig. 1. They exist in both defocusing [Figs. 1(a) and 1(b)] and focusing [Fig. 1(c)] media. Such solitons with $w_r, w_i \neq 0$ are characterized by complex internal energy flows, since gain in the system must be integrally compensated by losses. Physically, the truncation of potential creates negative surface defect (the area of surface channel is decreased in comparison with areas of other channels), that leads to beam expulsion into the depth of potential. At the same time, the truncation is performed in such way, that imaginary part $p_i R_i$ of potential yields gain in the surface channel stimulating near-surface localization in combination with nonlinearity. Near-surface gain can be partially compensated by uniform losses $\gamma$ and stable balance becomes possible between two competing effects mentioned above. While the field of soliton in focusing medium does not change its sign, in defocusing medium solitons feature oscillating tails indicating on the fact that such states are formed due to Bragg reflection from periodic structure. The parameters of solitons are fully determined by the strength of losses $\gamma$. The energy flow $U = \int_{-\infty}^{\infty}(w_r^2 + w_i^2)d\eta$ of soliton in focusing medium monotonically decreases with decrease of $\gamma$ [Fig. 2(a)]. When losses become too strong surface solitons cease to exist - in lower cutoff $\gamma = \gamma_{\text{low}}$ the tangential line to dependence $U(\gamma)$ becomes vertical. Increase of $\gamma$ is accompanied by the growth of peak amplitude and progressive light localization in the surface channel. In defocusing medium $U$ also grows with increase of $\gamma$ but in addition to lower cutoff there exists also upper cutoff $\gamma_{\text{upp}}$ [Fig. 2(a)]. In upper cutoff $b$ value for surface soliton approaches zero [Fig. 2(b)] that is accompanied by drastic expansion of soliton into uniform medium [Fig. 1(b)]. In defocusing medium $b$ decreases with increase of $\gamma$, in contrast to focusing medium. While in focusing medium weakest localization is achieved at $\gamma \to \gamma_{\text{low}}$, in defocusing medium low-energy solitons may be very well localized. The localization at $\gamma \to \gamma_{\text{low}}$ is determined by the amplitude $p_i$ of imaginary part of potential. When $p_i$ is small soliton drastically expands into the depth of potential for any $\gamma$, while for large $p_i$ values the solitons are well localized at $\eta > 0$ (although they may expand into uniform medium). Fig. 2(a) reveals a striking difference between solitons in focusing and defocusing media: in focusing medium stable solitons exist above certain energy flow threshold, while in defocusing medium they exist even for $U \to 0$. The latter fact indicates on the existence of linear guided mode, i.e. on possibility of "gain-guiding" modified by Bragg reflection from the lattice for unique value of $\gamma$. Lower cutoffs $\gamma_{\text{low}}$ for soliton existence almost coincide in focusing and defocusing media and they monotonically decrease with increase of $p_i$ [Fig. 2(c)].

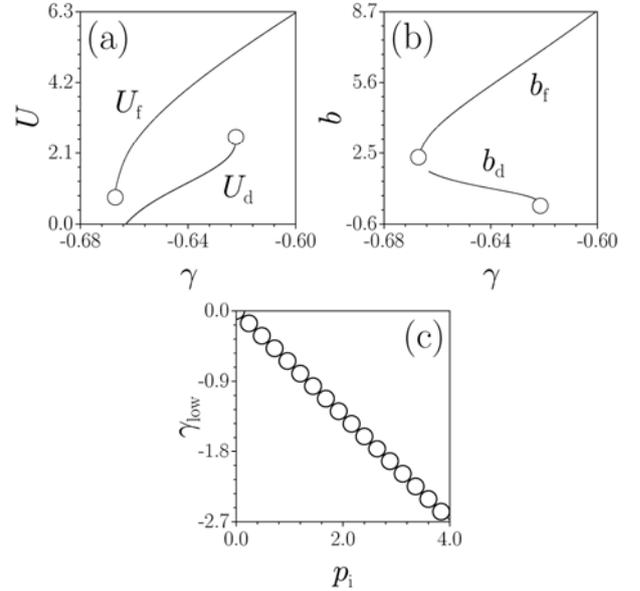

Fig. 2. Energy flow (a) and propagation constant (b) of surface soliton vs $\gamma$ in focusing (curves labeled with $U_f, b_f$) and defocusing (curves labeled $U_d, b_d$) media at $p_i = 1$. Circles correspond to solitons in Figs. 1(a),(c). (c) Lower border of existence domain for surface solitons in focusing medium as a function of $p_i$.

The central issue is the stability of surface solitons (Fig. 3). Stability analysis indicates that in focusing medium surface solitons can be stable in the domain adjacent to cutoff, i.e. for $\gamma_{\text{crl}} \leq \gamma \leq \gamma_{\text{cru}}$. Stability domain is depicted in Fig. 3(a) on the plane $(p_i, \gamma)$, where we subtracted cutoff $\gamma_{\text{low}}$ from $\gamma$ value for convenience. If losses are too weak, the solitons become unstable [a typical dependence $\delta(\gamma)$ is shown in Fig. 3(c)]. At small and moderate $p_i$ values the lower boundary of stability domain $\gamma_{\text{crl}}$ coincides with cutoff $\gamma_{\text{low}}$. At high $p_i$ values the instability domain appears also close to cutoff and for $p_i \approx 3.24$ [this value is indicated by circle in Fig. 3(a)] the entire soliton family becomes unstable. Although in terms of $\gamma$ the stability domain is narrow, the soliton's energy flow within this domain varies dramatically. In defocusing medium the entire soliton family is usually stable, i.e. stability domain is given by $\gamma_{\text{low}} \leq \gamma \leq \gamma_{\text{upp}}$ [Fig. 3(b)]. Only when $p_i \to 0$ a narrow instability domain appears near upper cutoff. The domain of stability slowly shrinks with increase of $p_i$.

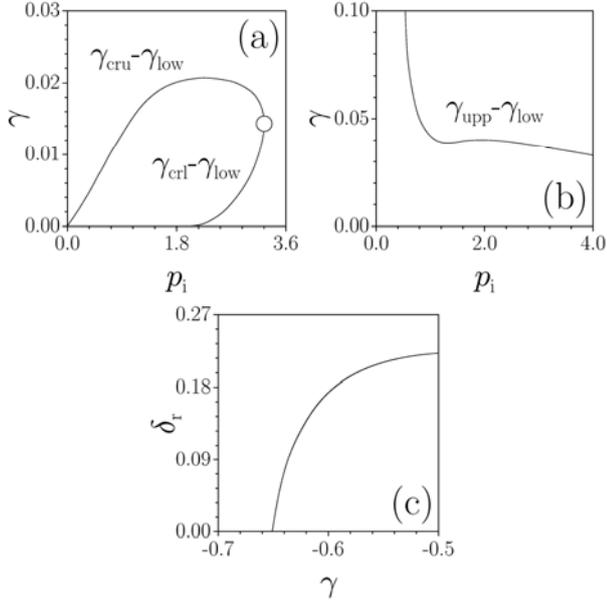

Fig. 3. Domains of stability for surface solitons in focusing (a) and defocusing (b) medium on the plane $(p_i, \gamma)$. In focusing medium solitons are stable in the area between curves $\gamma_{\text{cru}} - \gamma_{\text{low}}$ and $\gamma_{\text{crl}} - \gamma_{\text{low}}$, while in defocusing medium solitons are stable in the area below curve $\gamma_{\text{upp}} - \gamma_{\text{low}}$. (c) Real part of perturbation growth rate versus $\gamma$ for soliton in focusing medium at $p_i = 1$.

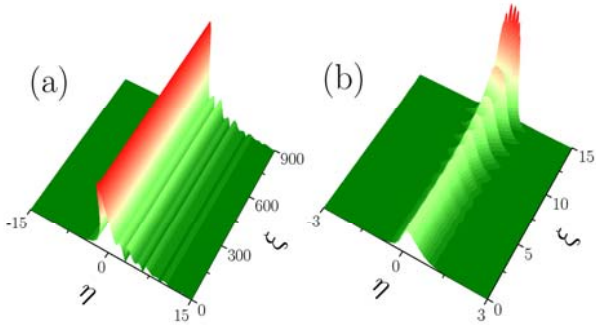

Fig. 4. (Colour online) Stable propagation of surface soliton in defocusing medium at $\gamma = -0.3249$, $p_i = 0.6$ (a) and growth of peak amplitude of unstable soliton in focusing medium at $\gamma = -1.2030$, $p_i = 2.0$ (b).

The example of stable propagation of perturbed surface soliton in defocusing medium is shown in Fig. 4(a). The field of such soliton considerably penetrates into the depth of potential. Despite strong perturbation the soliton retains its structure over indefinitely long distances. Because stable solutions reported here are attractors, they can be excited with various input beams even if they shifted into uniform medium or into the depth of potential. The perturbed solitons belonging to unstable branches are destroyed upon propagation. Usually the development of instability results in progressively growing oscillations of soliton center and rapid growth of peak amplitude [see example for focusing medium in Fig. 4(b)].

Summarizing, we showed that truncated periodic complex potentials with homogeneous losses can support stable surface solitons in both focusing and defocusing media. These results can be extended to other physical settings and geometries, including two-dimensional complex potentials where similar stable states may form.

This work was supported by the National Natural Science Foundation of China (Grant No. 11174061) and the Guangdong Province Natural Science Foundation of China (Grant No. S2011010005471).